\documentstyle[11pt]{article}
\makeatletter%
\def\nottoobig#1{{\hbox{$\left#1\vcenter to1.111\ht\strutbox{}\right.\n@space$}}}
\makeatother%

\makeatother%

\makeatletter%

\newcount\hour  \newcount\minutes  \hour=\time  \divide\hour by 60
\minutes=\hour  \multiply\minutes by -60  \advance\minutes by \time
\def\mmmddyyyy{\ifcase\month\or Jan\or Feb\or Mar\or Apr\or May\or Jun\or Jul\or
  Aug\or Sep\or Oct\or Nov\or Dec\fi \space\number\day, \number\year}
\def\hhmm{\ifnum\hour<10 0\fi\number\hour :%
  \ifnum\minutes<10 0\fi\number\minutes}
\def\Draft{{\it Draft of \mmmddyyyy}}

\def\ps@jtsheadings{%
\def\@oddhead{\it\rightmark\hfil\rm\thepage}%
\def\@oddfoot{\hfil\Draft}%
\if@twoside%
\def\@evenhead{\rm\thepage\hfil\it\leftmark}%
\def\@evenfoot{\Draft\hfil}%
\else
\let\@evenhead\@oddhead%
\let\@evenfoot\@oddfoot%
\fi%
}
\def\ps@jtsplain{%
\def\@oddhead{\hfil\Draft}%
\def\@oddfoot{\hfil\rm\thepage\hfil}%
\let\@evenfoot\@oddfoot%
\if@twoside \def\@evenhead{\Draft\hfil} \else \let\@evenhead\@oddhead \fi
}

\def\chaptermark#1{\markboth{\thechapter.\ #1}{\thechapter.\ #1}}%
\def\sectionmark#1{\markright{\thesection.\ #1}}

\def\section{\@startsection {section}{1}{\z@}
    {3.5ex plus1ex minus.2ex}{2.3ex plus.2ex}{\Large\bf}}
\def\subsection{\@startsection{subsection}{2}{\z@}
    {3.25ex plus1ex minus.2ex}{1.5ex plus.2ex}{\large\bf}}
\def\subsubsection{\@startsection{subsubsection}{3}{\z@}
    {3.25ex plus1ex minus.2ex}{1.5ex plus.2ex}{\normalsize\bf}}
\def\paragraph{\@startsection{paragraph}{4}{\z@}
    {3.25ex plus1ex minus.2ex}{1em}{\normalsize\bf}}
\def\subparagraph{\@startsection{subparagraph}{4}{\parindent}
    {3.25ex plus1ex minus.2ex}{1em}{\normalsize\bf}}

\makeatother%

\makeatletter \@beginparpenalty=10000 \makeatother

\def\underl#1 {\leavevmode\let\first=\relax\underli #1 }
\def\underli#1 {\ifx&#1\let\next=\relax\unskip
                \else\let\next=\underli\first\ulinebox{#1}\fi\let\first=\undersp\next}
\def\undersp{\penalty50\ulinebox{\space}\penalty50}
\def\ulinebox#1{\vtop{\hbox{\strut#1}\hrule}}%
\def\unice#1 {\underl #1 & }
\def\desclabel#1{\bf #1\hfil}
\def\desc{\list{}{%
\labelwidth=\leftmargin
\advance \labelwidth by -\labelsep
\let \makelabel=\desclabel}}

\makeatletter %

\newlength{\filength}
\newsavebox{\gcbox}
\sbox{\gcbox}{\framebox[\filength]{\rule{0ex}{2ex}}}

\newlength{\leftjustindent}
\newlength{\@leftjustindent}
\def\leftjust{\let\\\@leftjustcr\let\end\@endleftjust
  \addtolength{\@leftjustindent}{\leftjustindent}
  \vcenter\bgroup
  \halign\bgroup
    \hbox to\displaywidth{
      \rule{\@leftjustindent}{0ex}$\displaystyle##$\hfill
      }\crcr
}
\def\endleftjust{\crcr\egroup\egroup\endgroup}
\def\@endleftjust#1{\crcr\egroup\egroup\@checkend{#1}\endgroup}
\def\@leftjustcr{\crcr}

\newcommand{\redttnp}[1]{ {  {\rm R}_{{#1}{\scriptsize\mbox{-tt}}}^{p}({\np}) }    }
\newcommand{\redttbhk}[1]{ {  {\rm R}_{{#1}{\scriptsize\mbox{-tt}}}^{p}({\bh_k}) }    }
\newcommand{\redttbhj}[1]{ {  {\rm R}_{{#1}{\scriptsize\mbox{-tt}}}^{p}({\bh_j}) }    }
\newcommand{\redttbhl}[1]{ {  {\rm R}_{{#1}{\scriptsize\mbox{-tt}}}^{p}({\bh_l}) }    }

\newtheorem{theorem}{Theorem}[section]
\newtheorem{corollary}[theorem]{Corollary}
\def\Proof{\par\medskip\noindent{\bf Proof}\\\mbox{\rule{\parindent}{0in}}}
\newcommand{\qedblob}{\mbox{\rule[-1.5pt]{5pt}{10.5pt}}}
\def\literalqed{{\ \nolinebreak\hfill\mbox{\qedblob\quad}}}

\def\qed{\literalqed}

\newtheorem{lemma}[theorem]{Lemma}

\newtheorem{fact}[theorem]{Fact}

\newcommand{\singlespacing}{\let\CS=
\@currsize\renewcommand{\baselinestretch}{1}\tiny\CS}
\newcommand{\singlespacingplus}{\let\CS=
\@currsize\renewcommand{\baselinestretch}{1.25}\tiny\CS}
\newcommand{\doublespacing}{\let\CS=
\@currsize\renewcommand{\baselinestretch}{1.75}\tiny\CS}
\newcommand{\draftspacing}{\let\CS=
\@currsize\renewcommand{\baselinestretch}{2.0}\tiny\CS}

\makeatother%

\mathcode`\0="0030      %
\mathcode`\1="0031
\mathcode`\2="0032
\mathcode`\3="0033
\mathcode`\4="0034
\mathcode`\5="0035
\mathcode`\6="0036
\mathcode`\7="0037
\mathcode`\8="0038
\mathcode`\9="0039
\makeatletter
\clubpenalty=\@highpenalty
\widowpenalty=\@highpenalty
\makeatother

\makeatletter
\newcommand{\niceonespacing}{\let\CS=\@currsize\renewcommand{\baselinestretch}{1.1}\tiny\CS}\newcommand{\nicetwospacing}{\let\CS=\@currsize\renewcommand{\baselinestretch}{1.2}\tiny\CS}
\newcommand{\nicethreespacing}{\let\CS=\@currsize\renewcommand{\baselinestretch}{1.3}\tiny\CS}
\newcommand{\singlespacingplusplus}{\let\CS=\@currsize\renewcommand{\baselinestretch}{1.35}\tiny\CS}
\newcommand{\nicefourspacing}{\let\CS=\@currsize\renewcommand{\baselinestretch}{1.4}\tiny\CS}
\newcommand{\nicefivespacing}{\let\CS=\@currsize\renewcommand{\baselinestretch}{1.5}\tiny\CS}
\newcommand{\nicesixpacing}{\let\CS=\@currsize\renewcommand{\baselinestretch}{1.6}\tiny\CS}
\makeatother

\makeatletter
\def\@cite#1#2{[#1\if@tempswa , #2\fi]}
\makeatother

\makeatletter
\def\@citex[#1]#2{\if@filesw\immediate\write\@auxout{\string\citation{#2}}\fi
  \def\@citea{}\@cite{\@for\@citeb:=#2\do
    {\@citea\def\@citea{,\linebreak[0]}\@ifundefined
       {b@\@citeb}{{\bf ?}\@warning
       {Citation `\@citeb' on page \thepage \space undefined}}%
\hbox{\csname b@\@citeb\endcsname}}}{#1}}
\makeatother

\makeatletter
\def\ps@thesis{\def\@oddhead{\hfil\rm\thepage\hfil}\def\@oddfoot{}\def\@evenhead{\hfil\rm\thepage\hfil}\def\@evenfoot{}\def\chaptermark##1{}\def\sectionmark##1{}}
\makeatother

\makeatletter
\def\foobarpt{\textfont\z@\tenrm 
  \scriptfont\z@\ninrm \scriptscriptfont\z@\sevrm
\textfont\@ne\tenmi \scriptfont\@ne\ninmi \scriptscriptfont\@ne\sevmi
\textfont\tw@\tensy \scriptfont\tw@\ninsy \scriptscriptfont\tw@\sevsy
\textfont\thr@@\tenex \scriptfont\thr@@\tenex \scriptscriptfont\thr@@\tenex
\def\unboldmath{\everymath{}\everydisplay{}\@nomath\unboldmath
          \textfont\@ne\tenmi 
          \textfont\tw@\tensy \textfont\lyfam\tenly
          \@boldfalse}\@boldfalse
\def\boldmath{\@ifundefined{tenmib}{\global\font\tenmib\@mbi\@magscale1\global
        \font\tensyb\@mbsy \@magscale1\global\font
         \tenlyb\@lasyb\@magscale1\relax\@addfontinfo\@xiipt
              {\def\boldmath{\everymath
                {\mit}\everydisplay{\mit}\@prtct\@nomathbold
                \textfont\@ne\tenmib \textfont\tw@\tensyb 
                \textfont\lyfam\tenlyb\@prtct\@boldtrue}}}{}\@xiipt\boldmath}%
\def\prm{\fam\z@\tenrm}%
\def\pit{\fam\itfam\tenit}\textfont\itfam\tenit \scriptfont\itfam\ninit
   \scriptscriptfont\itfam\sevit
\def\psl{\fam\slfam\tensl}\textfont\slfam\tensl 
     \scriptfont\slfam\tensl \scriptscriptfont\slfam\tensl
\def\pbf{\fam\bffam\tenbf}\textfont\bffam\tenbf 
   \scriptfont\bffam\ninbf \scriptscriptfont\bffam\ninbf 
\def\ptt{\fam\ttfam\tentt}\textfont\ttfam\tentt
   \scriptfont\ttfam\nintt \scriptscriptfont\ttfam\nintt 
\def\psf{\fam\sffam\tensf}\textfont\sffam\tensf
    \scriptfont\sffam\tensf \scriptscriptfont\sffam\tensf
\def\psc{\@getfont\psc\scfam\@xiipt{\@mcsc\@magscale1}}%
\def\ly{\fam\lyfam\tenly}\textfont\lyfam\tenly 
   \scriptfont\lyfam\ninly \scriptscriptfont\lyfam\sevly
 \@setstrut \rm}

\makeatother

\newcommand{\p}{{\rm P}}

\newcommand{\np}{{\rm NP}}

\newcommand{\bh}{{\rm BH}}
\newcommand{\cobh}{{\rm coBH}}
\newcommand{\pcalonetwo}{  {\rm P}^{ {\cal C}_1 [1] : {\cal C}_2 [1]}}
\newcommand{\pcalonetwothree}{  {\rm P}^{ {\cal C}_1 [1] : {\cal C}_2 [1] ,
{\cal C}_3 [1]}}

\newcommand{\pjk}{  {\rm P}^{ {\rm BH}_j [1] : {\rm BH}_k [1]}}

\newcommand{\pttjttkttl}{  {\rm P}^{ \redttbhj{1} [1] : \redttbhk{1} [1] , \redttbhl{1} [1]}}
\newcommand{\pjkl}{  {\rm P}^{ {\rm BH}_j [1] : {\rm BH}_k [1] , {\rm BH}_l [1]}}
\newcommand{\pjprimekprime}{  {\rm P}^{ {\rm BH}_{j'} [1] : {\rm BH}_{k'} [1]}}
\newcommand{\pkj}{  {\rm P}^{ {\rm BH}_k [1] : {\rm BH}_j [1]}}

\newcommand{\npnp}{{\np^{\rm NP}}}

\newcommand{\sigmak}{{\Sigma_k^p}}
\newcommand{\pik}{{\Pi_k^p}}

\newcommand{\ph}{{\rm PH}}

\newcommand{\calc}{\mbox{$\cal C$}}
\newcommand{\calone}{\mbox{${\cal C}_1$}}
\newcommand{\caltwo}{\mbox{${\cal C}_2$}}
\newcommand{\calthree}{\mbox{${\cal C}_3$}}

\newcommand{\condition}{\,\nottoobig{|}\:}
\def\land{{\; \wedge \;}}
\title{Query Order}

\author{Lane A. Hemaspaandra\protect\thanks{
Department of Computer Science, 
University of Rochester
Rochester, NY 14627, USA\@.  Supported in part 
by grants NSF-CCR-9322513
and NSF-INT-9513368/DAAD-315-PRO-fo-ab.  
Work done in part while 
visiting 
Friedrich-Schiller-Universit\"at Jena.
\mbox{lane@cs.rochester.edu}.}
\and Harald Hempel\protect\thanks{
Institut~f\"ur Informatik,
Friedrich-Schiller-Universit\"at Jena,
07740 Jena, Germany.  Supported in part 
by grant NSF-INT-9513368/DAAD-315-PRO-fo-ab.
\mbox{hempel@informatik.uni-jena.de}.}
\and Gerd Wechsung\protect\thanks{
Institut f\"ur Informatik,
Friedrich-Schiller-Universit\"at Jena,
07740 Jena, Germany.  Supported in part 
by grant 
NSF-INT-9513368/\protect\linebreak[2]DAAD-315-PRO-fo-ab. \protect\linebreak[3]
\mbox{wechsung@informatik.uni-jena.de}.}
}

\makeatletter
\def\@listI{\leftmargin\leftmargini \parsep 4.5pt plus 1pt minus 1pt\topsep
6pt plus 2pt minus 2pt \itemsep  2pt plus 2pt minus 1pt}

\let\@listi\@listI
\@listi
\makeatother

\begin{document}

\bibliographystyle{alpha}

\setcounter{page}{1}

{\singlespacing\maketitle}

\thispagestyle{plain}
\markboth{L.A.~HEMASPAANDRA, H.~HEMPEL, AND
G.~WECHSUNG}{QUERY ORDER}

\begin{abstract}
We study the effect of query order on
computational power, and show that $\pjk$---the languages 
computable via a polynomial-time machine given 
one query to the $j$th level of the 
boolean hierarchy followed by one query to the $k$th 
level of the boolean hierarchy---equals $\redttnp{j+2k-1}$
if $j$ is even and $k$ is odd, and equals
$\redttnp{j+2k}$ otherwise.  
Thus, unless the polynomial hierarchy
collapses, it holds that for each $1\leq j \leq k$:
$\pjk = \pkj \iff 
(j=k) \lor (j\mbox{ is even}\, \land
k=j+1)$. 
We extend our analysis to apply to more general query classes.
\end{abstract}
\setcounter{page}{1}
\sloppy
\section{Introduction}\label{s:intr}

This paper studies the importance of query order.
Everyone knows that it makes more sense to first look up in your 
on-line datebook the date of the yearly computer science conference 
and then phone your travel agent to get tickets, as opposed to first 
phoning your travel agent (without knowing the date) and then 
consulting your on-line datebook to find the date. 
In real life, order matters.

This paper seeks to determine---for the first time to the best of our 
knowledge---whether one's everyday-life intuition that order matters 
carries over to complexity theory.

In particular, for classes $\calone$ and
$\caltwo$ from the boolean 
hierarchy~\cite{cai-gun-har-hem-sew-wag-wec:j:bh1,cai-gun-har-hem-sew-wag-wec:j:bh2},
we ask whether one question to $\calone$ followed by one question 
to $\caltwo$ is more powerful than one question to $\caltwo$ followed 
by one question to $\calone$.  That is, we seek the 
relative powers of the classes $
\p^{ {\cal C}_1[1]: {\cal C}_2[1]}$ and 
$
\p^{ {\cal C}_2[1]: {\cal C}_1[1]}$.

As is 
standard~\cite{lad-lyn-sel:j:com}, 
for any constant $m$ we say $A$ is $m$-truth-table reducible to $B$ 
($A \leq_{m\hbox{-}{\rm tt}}^p B$) 
if there is a polynomial-time computable function 
that, on each input $x$, computes both (a) $m$ strings 
$x_1,x_2,\cdots,x_m$ and (b) a predicate, $\alpha$, of $m$ boolean 
variables, such that $x_1,x_2,\cdots,x_m$ and $\alpha$ satisfy:
$$x \in A \iff \alpha(\chi_B(x_1),\chi_B(x_2),\cdots ,\chi_B(x_m)),$$
where $\chi_B$ denotes the characteristic function of $B$.
For any $a$ and $b$ for which $\leq_a^b$ is defined and 
for any class $\calc$, let 
${\rm R}_a^b(\calc)=\{L \condition (\exists C \in \calc)[L \leq_a^b C]\}$.

We prove via the mind change  technique 
that, for $j,k\geq 1$:
$$
\pjk = \left\{
\begin{array}{l}
\protect\redttnp{j+2k-1} \mbox{~~~~~~~~~~~~~~~~~~~~~if $j$ is even and $k$ is odd}\\
\protect\redttnp{j+2k}    \mbox{~~~~~~~~~~~~~~~~~~~~~~~~otherwise.}
\end{array}
\right.$$
Informally, this says that the second query counts more 
towards the power of the 
class than the first query does.  In particular, 
assuming that the polynomial hierarchy does not collapse, we have 
that if $1\leq j \leq k$ then $\pjk$ and $\pkj$ differ 
unless $(j=k) \lor (j$ is even$\,\land k = j+1)$.  

The interesting case is the strength of $\pjk$ when $j$ is even and
$k$ is odd.  In some sense, $j+2k$ NP questions underpin this class.
However, 
by arguing that a certain underlying graph must contain an
odd cycle, 
we show that one can always make do with $j+2k-1$ queries.
We generalize our results to apply broadly to classes with tree-like 
query structure.

The work of this paper (especially Theorem~\ref{t:main}), which 
first appeared in~\cite{hem-hem-wec:tOutBySICOMP:query-order-bh},
should be compared with the independent work of Agrawal, Beigel, and
Thierauf,
which first appeared 
in~\cite{agr-bei-thi:tOutByFSTTCSConf:modulo-information}.
In particular,
let $\p^{{\rm BH}_j [1]:{\rm BH}_k [1]_+}$ denote the class of languages 
recognized by some polynomial-time machine making one query to a 
${\rm BH}_j$ oracle 
followed by one query to a ${\rm BH}_k$ oracle and accepting if and only if 
the second query is answered ``yes.''
Agrawal, Beigel, and Thierauf prove (using different notation):\quad 
$$
\p^{{\rm BH}_j [1]:{\rm BH}_k [1]_+} = \left\{
\begin{array}{l}
\protect{\rm BH}_{j+2k-1} \mbox{~~~~~~~~~~~~~~~~~~~~~if $j \not\equiv k \pmod{2}$}\\
\protect{\rm BH}_{j+2k}    \mbox{~~~~~~~~~~~~~~~~~~~~~~~~otherwise.}
\end{array}
\right.$$

Note that this 
result is incomparable with the results of Theorem~\ref{t:main}, as 
their result deals with a different 
and seemingly more restrictive acceptance mechanism.  Some 
insight into the degree of restrictiveness of their
acceptance mechanism, and its relationship to ours, is given by
the following 
claim (Corollary~\ref{c:rel-r-h}), 
which follows immediately from 
Theorem~5.7 and Lemma~5.9 of~\cite{agr-bei-thi:tOutByFSTTCSConf:modulo-information}
and Theorem~\ref{t:main} of the present paper:
$${\rm R}^p_{1{\scriptsize\mbox{-tt}}}(\p^{{\rm BH}_j [1]:{\rm BH}_k [1]_+}) = 
\left\{
\begin{array}{l}
\protect{\p^{{\rm BH}_{j-1} [1]:{\rm BH}_k [1]}} 
\mbox{~~~~~~~~~~~~~~~~~~~~~if $j$ is odd and $k$ is even}\\
\protect\pjk \mbox{~~~~~~~~~~~~~~~~~~~~~~~~otherwise.}
\end{array}
\right.$$

\section{Preliminaries}\label{section:prelim}

For standard notions not defined here, we refer
the reader to any computational complexity
textbook, 
e.g.,~\cite{bov-cre:b:complexity,bal-dia-gab:b:sctI-2nd-ed,pap:b:complexity}.
Let $\chi_A$,
$m$-truth-table reducibility (``$\leq_{m\hbox{-}{\rm tt}}^p$''),
${\rm R}_a^b(\calc)$, and 
$\p^{{\rm BH}_j [1]:{\rm BH}_k [1]_+}$ be as defined in
Section~\ref{s:intr}.

The boolean 
hierarchy~\cite{cai-gun-har-hem-sew-wag-wec:j:bh1,%
cai-gun-har-hem-sew-wag-wec:j:bh2}
is defined as follows, where $\calone  \ominus \caltwo = 
\{L\condition (\exists A\in \calone)(\exists B \in \caltwo)[L=A-B]\}$.
$$\bh_1 = \np,$$
$$\bh_k = \np \ominus \bh_{k-1}, ~\mbox{for }k>1,$$
$$\cobh_k= \{ L \condition \overline{L}\in \bh_k\}, ~\mbox{for } k > 0,\mbox{ and}$$
$$\bh= \bigcup_{i\geq 1} \bh_i.$$
The boolean hierarchy has been intensely investigated,
and quite a bit has been learned about its structure
(see, e.g., \cite{cai-gun-har-hem-sew-wag-wec:j:bh1,%
cai-gun-har-hem-sew-wag-wec:j:bh2,%
cai:c:boolean-1,koe-sch-wag:j:diff,kad:joutdatedbychangkadin:bh,%
wag:j:bounded,cha-kad:j:closer,%
bei:j:bounded-queries,bei-cha-ogi:j:difference-hierarchies}).
Recently, various results have also been developed regarding 
boolean hierarchies over classes 
other than NP~\cite{bru-jos-you:j:strong,cha:thesis:boolean,%
bei-cha-ogi:j:difference-hierarchies,hem-rot:j:boolean}.

For any language classes $\calone$ and $\caltwo$, define
$\pcalonetwo$ to be the class of languages accepted by polynomial-time
machines making one query to a $\calone$ oracle followed by one query 
to a $\caltwo$ oracle.  
For any language classes $\calone$, $\caltwo$, and $\calthree$, define
$\pcalonetwothree$ to be the class of languages accepted by polynomial-time
machines making one query to a $\calone$ oracle followed in the case of a ``no'' answer 
to this first query by one query to a $\caltwo$ oracle, and in 
the case of a ``yes'' answer to the first query by one query to a 
$\calthree$ oracle.

\section{The Importance of Query Order} \label{section:order}

We ask whether the order of queries matters.  We will study this 
in the setting of the boolean hierarchy.  In particular, 
does $\pjk$ equal $\pkj$, or are they incomparable, or does one 
strictly contain the other?  

For clarity of
presentation, we will in this section handle classes only of the form 
$\pjk$. 
We show that 
for no $j$, $k$, $j'$, and $k'$ are $\pjk$ and 
$\pjprimekprime$ incomparable. 
In Section~\ref{section:general} we will handle the more general case of classes 
of the form $\pjkl$, and even classes with a more complicated tree-like query 
structure. 

Indeed, we show in this section that in almost all 
cases,
$\pjk$ is so powerful
that it can do anything that can be done with $j+2k$ truth-table 
queries to NP\@.  Since, based on the answer to the first $\bh_j$
query, there are two possible $\bh_k$ queries that might follow, $j+2k$ is 
exactly the number of queries asked in a brute force truth-table 
simulation of $\pjk$.  Thus, our result shows that (in almost
all cases) the power of the class is 
not reduced by the nonlinear structure of 
the $j+2k$ queries underlying $\pjk$---that is,
the power is 
not reduced 
by the fact that in any given run only $j+k$ underlying NP queries will
be even implicitly asked (via the $\bh_j$ query and the one 
asked $\bh_k$ query).  We say ``in almost all cases'' as if $j$ is even
and $k$ is odd, we prove there is a power reduction of exactly one 
level.

All the results of the previous paragraph follow from a
general characterization that we prove.  For $j,k \geq 1$:
$$
\pjk = \left\{
\begin{array}{l}
\protect\redttnp{j+2k-1} \mbox{~~~~~~~~~~~~~~~~~~~~~if $j$ is even and $k$ is odd}\\
\protect\redttnp{j+2k}    \mbox{~~~~~~~~~~~~~~~~~~~~~~~~otherwise.}
\end{array}
\right.$$
Our proof employs the mind change technique,
which predates complexity theory. 
In particular we show that $\pjk$ has at most $j+2k$ ($j+2k-1$ if $j$ is even and 
$k$ is odd) mind changes, and that ${\rm BH}_{j+2k}$ (${\rm BH}_{j+2k-1}$ if $j$ 
is even and $k$ is odd) is contained in $\pjk$.

  The mind change technique or equivalent manipulation
was applied to complexity theory in each of the 
early papers on the boolean hierarchy, including 
the work of 
Cai et 
al.~(\cite{cai-gun-har-hem-sew-wag-wec:j:bh1}, see 
also~\cite{wec:c:bh:ormaybe:wech:only:is:right,cai-hem:c:boolean}),
K\"obler et al.~\cite{koe-sch-wag:j:diff}, and 
Beigel~\cite{bei:j:bounded-queries}.  
These papers use mind changes for a number of purposes.  Most 
crucially they use the maximum number of mind changes 
(what a mind change
is will soon be made clear) of a class
as an upper bound that can be used to prove that the class is 
contained in some other class.  
In the other direction,
they also use the number of mind changes
that certain classes---especially the classes of 
the boolean hierarchy due to their normal form
as nested subtractions of telescoping
sets~\cite{cai-gun-har-hem-sew-wag-wec:j:bh1}---possess to 
show that they can simulate other classes.
Even for classes that have the same number of mind changes,
relativized separations are obtained via showing that the 
mind changes are of different character (mind change sequences are 
of two types, depending on whether they start with acceptance or 
rejection).
The technique has also 
proven useful in many other 
more recent papers, e.g.,~\cite{cha-kad:j:closer,cha:thesis:boolean,bei-cha-ogi:j:difference-hierarchies}.

To make clear the basic nature of mind change arguments, in a simple
form, we give an example.  We informally argue that each set that is
$k$-truth-table reducible to NP is in fact in $ \redttbhk{1}$.  

\begin{lemma}\label{l:tt}
For every $k \geq 1$, $\redttnp{k}=\redttbhk{1}$.
\end{lemma}

\noindent{\bf Proof} \quad
This fact (stated slightly differently) is due to K\"obler et
al.~\cite{koe-sch-wag:j:diff}, and the proof flavor presented 
here is most akin to the 
approach of Beigel~\cite{bei:j:bounded-queries}.  
Consider a $k$-truth-table reduction
to an NP set, $F$\@.  Let $L$ be the language accepted by the
$k$-truth-table reduction to $F$\@.  Consider some input $x$ and
without loss of generality assume $k$ queries are generated.  Let us
suppose for the moment that the reduction rejects when all $k$ queries
receive the answer ``no.''  Consider the $k$-dimensional hypercube
such that one dimension is associated with each 
query (0 in that dimension means the query is answered no and 
1 means it is answered yes).  
So the origin is associated with all queries being
getting the answer no and the point (1,1,...,1) is associated with all
queries getting the answer yes.  Now, also label each vertex with
either A (accept) or R (reject) 
based on what the truth-table would do given the answers
represented by that vertex.  So under our supposition, the origin has
the label R\@.  Finally, label each vertex with an integer as follows.
Label the origin with 0.  Inductively label each remaining vertex with the
{\em maximum\/} integer induced by the vertices that immediately
precede it (i.e., those that are the same as it except one yes answer
has been changed to a no answer).  A preceding vertex $v$ with integer
label $i$ induces in a successor $v'$ the integer $i+1$ if $v$ and
$v'$ have different A/R labels, and $i$ if they have the same
label.  Note
that vertices given even labels correspond to rejection and those
given odd labels correspond to acceptance.  Informally, a mind change
is just changing one or more strings from no to yes in a way that 
moves us from a vertex labeled $i$ to one labeled $i+1$.
For $1\leq i \leq k$, let
$B_i$ be the NP set that accepts $x$ if (in the queries/labeling
generated by the action of the truth-table on input $x$) for some
vertex $v$ labeled $i$ all the queries $v$ claims are yes
are indeed in the NP set $F$.
Note
that $B_1 \supseteq B_2 \supseteq B_3 \supseteq \cdots$, as if a node
labeled $v$ is in $B_j$, $j \geq 2$, then certainly its predecessor
node with label $j-1$ must be in $B_{j-1}$, as that predecessor
represents a subset of the strings $v$ represents.  But now note that
$L$ is exactly $B_1 - ( B_2 - ( B_3 - ( \cdots - (B_{k-1} - B_k)
\cdots )))$.  Why?  Let the vertex $w$ (say with integer label $i_w$)
represent the true answers to the queries.  Note that by construction,
$x \in B_{q}$ for all $q \leq i_w$ but $x \not\in
B_q$ for any $q> i_w$.  As the $B_i$ were alternating in terms of
representing acceptance and rejection, and given the format $B_1 - (
B_2 - ( B_3 - ( \cdots - (B_{k-1} - B_k) \cdots )))$, the set $B_1 - (
B_2 - ( B_3 - ( \cdots - (B_{k-1} - B_k) \cdots )))$, will do exactly
what $B_{i_w}$ represents, namely, the action on the correct answers.
Thus, we have just given a proof that a $k$-truth-table reduction that
rejects whenever all answers are no can be simulated by a set in
$\bh_k$.  Of course, one cannot validly assume that the reduction
rejects whenever all answers are no.  But it is not hard to see
(analogously to the above) that the case of inputs where the reduction
accepts when all answers are no can (analogously to the above) be
handled via the complement of a $\bh_k$ set, and that (since what the
truth-table reduction does when all answers are no is itself
polynomial-time computable) via a set in $\redttbhk{1}$ we can accept
an arbitrary set in $\redttnp{k}$.  Of course, it is clear by brute
force simulation that $\redttbhk{1} \subseteq \redttnp{k}$, and so it
holds that $\redttbhk{1} = \redttnp{k}$.~\qed

What actually is being shown above is that $\redttbhk{1}$ can handle 
$k$ appropriately structured mind changes, starting either from reject 
or accept.  In the following theorem, the crucial things we show 
are that (a)~$\pjk$ can simulate, starting at either accept or 
reject, $j+2k$ (respectively, $j+2k-1$) mind changes 
if $j$ is odd or $k$ is even (respectively, if $j$ is even and $k$
is odd), and (b)~for $j$ even and $k$ odd, $\pjk$ can {\em never\/} have
more than $j+2k-1$ mind changes.
We achieve~(b)~by examining the possible mind 
change flow of a $\pjk$ machine, $j$ even and $k$ odd, and 
showing that either a mind change is flagrantly wasted, or a certain
underlying graph has an odd length directed cycle (which thus is not 
two-colorable, and from this will lose one mind change).

Since our arguments in the proofs of this section use
paths in hypercubes, we will find useful
the concept of an ascending path in a hypercube.
Let $K=\{0,1\}^d$ be the $d$-dimensional hypercube. 
Then every path 
$p$ in $K$ can be described as a linear 
combination of unit vectors $u_1 , \ldots , u_d$, where $u_i$ is the 
$i$th unit vector. 
We call $p$ an ascending path in $K$ leading from $(0,0,\cdots,0)$ to $v$ 
if and only if it can be 
identified with a sum $$u_{i_1}+u_{i_2}+ \cdots +u_{i_n}$$ of 
distinct unit vectors $u_{\nu}$ such that the vertices of this path $p$ are
$$v_0=(0,\cdots,0), v_1=u_{i_1}, v_2=u_{i_1}+u_{i_2}, \cdots{}, 
v=u_{i_1}+u_{i_2}+ \cdots +u_{i_n} .$$
We will call this sum the description of $p$.
Note that the order of the $u$'s matters, as a permutation of the $u$'s 
results in another path.
We call $p$ an ascending path (without specifying starting point and 
endpoint) if $p$ 
is an ascending path leading from $(0,0,\cdots,0)$ to $(1,1,\cdots,1)$.

Before turning to results,
we will first
study the structure of ascending 
paths in labeled hypercubes 
and give some necessary definitions. Building upon them, we will then 
prove Lemma~\ref{l:mind}, which states that $\pjk$ can handle exactly 
$j+2k$ ($j+2k-1$ if $j$ is even and $k$ is odd) mind changes.

Let $M$ be a $\pjk$ machine with oracles $A\in\bh_j$ and $B\in\bh_k$ 
and let $x \in \Sigma^*$.
On input $x$, $M$ first makes a query $q_1(x)$ to $A$ and then if the 
answer to the first query was ``no'' asks query $q_2(x)$ to $B$ and 
if the answer to the first query was ``yes'' asks query $q_3(x)$ to 
$B$. 
Without loss of generality assume that on every input $x$ exactly two 
queries are asked.

Every set $C \in \bh_l$ can be written as the nested difference of 
sets $C_1,C_2,\cdots , C_l \in \np$ 
$$C=C_1-(C_2-(\cdots -(C_{l-1}-C_l)\cdots ))$$ 
and following Cai et al.~\cite{cai-gun-har-hem-sew-wag-wec:j:bh1} we 
even can assume that 
$$C_l \subseteq C_{l-1} \subseteq \cdots  \subseteq C_2 \subseteq C_1.$$
Hence a query ``$q \in C$?'' can certainly be solved via $l$ queries 
``$q \in C_1$?,'' ``$q \in C_2$?,'' $\cdots$ ,
``$q \in C_l$?''

In light of this comment, we let 
\begin{eqnarray*}
A=A_1-(A_2-(\cdots-(A_{j-1}-A_j)\cdots)) \quad &\mbox{where}& 
\quad A_i \in \np \quad 
\mbox{for} \quad  i=1,2,\cdots ,j\\ 
&&\quad \mbox{and} \quad A_j \subseteq \cdots \subseteq A_1,\mbox{ and}
\end{eqnarray*}
\begin{eqnarray*}
B=B_1-(B_2-(\cdots-(B_{k-1}-B_k)\cdots)) \quad &\mbox{where}& 
\quad B_i \in \np \quad 
\mbox{for} \quad i=1,2,\cdots ,k \\
&&\quad \mbox{and} \quad B_k \subseteq \cdots 
\subseteq B_1.
\end{eqnarray*}

For the sake of definiteness let us assume that the queries
$$q_1(x)\in A_1,\cdots,q_1(x)\in A_j,q_2(x)\in B_1,\cdots,q_2(x)\in 
B_k,q_3(x)\in B_1,\cdots,q_3(x)\in B_k$$
correspond in this order to the $j+2k$ dimensions of the $(j+2k)$-dimensional 
hypercube $H=\{0,1\}^{j+2k}$. More precisely, a vector $(a_1,\cdots,a_{j+2k})
 \in H$ is understood to consist of the answers to the above-mentioned 
queries, where 0 means ``no'' and 1 means ``yes.''

Since a query ``$ q \in C$?'' for some $C \in \bh_l$ and 
$C=C_1-(C_2-(\cdots (C_{l-1}-C_l)\cdots ))$ can be solved by evaluating
the answers to ``$q \in C_1$?,'' ``$q \in C_2$?,'' ..., ``$q \in C_l$?'' 
every node $v \in H$ gives us answers to ``$q_1(x) \in A$?'' (by evaluating
the first $j$ components of $v$), to ``$q_2(x) \in B$?'' (by evaluating 
the $k$ components of $v$ that immediately follow the first $j$
components of $v$) and to ``$q_3(x) \in B$?'' (by evaluating 
the last $k$ of $v$'s components). 
This gives us a labeling of all vertices of $H$.  We simply assign label 
A (Accept) to vertex $v \in H$ if $M^{A[1]:B[1]}(x)$ accepts if the 
answers to the two asked questions are as determined by $v$. If 
$M^{A[1]:B[1]}(x)$ rejects in this case we assign label R (Reject) to $v$.
 
So let $H_M(x)$ be the $(j+2k)$-dimensional hypercube labeled 
according to $M^{A[1]:B[1]}(x)$. 
The number of mind changes on an ascending path $p$ of $H_M(x)$ leading from 
$(0,0, \cdots,0)$ to a vertex $t$ is by definition the number of label 
changes when moving from $(0,0,\cdots,0)$ to $t$ along $p$. 
The number of mind changes of an internal node $v$ of $H_M(x)$ is the 
maximum number of mind changes on an ascending path leading from 
$(0,0, \cdots,0)$ to $v$.
And finally, the number of mind changes of a $\pjk$ machine $M$ is by 
definition the maximum number (we take the maximum over all $x \in \Sigma ^*$) 
of mind changes of the vertex $(1,1,\cdots,1)$ in $H_M(x)$;
in other words, this number is the maximum number of label changes 
on an ascending 
path in $H_M(x)$ for some $x \in \Sigma ^*$.

We say we lose a mind change (between two adjacent vertices $v_i$ 
and $v_{i+1}$) along an ascending path if when 
moving from $v_i$ to $v_{i+1}$ the machine does not change its acceptance 
behavior. 

One can easily verify the following fact:

\begin{fact}\label{f:2}
If $M$ is a $\pjk$ machine such that on input $x$ the acceptance 
behavior is 
independent of the answer to one or more of the two possible second 
queries (that is, if for at least one of the second queries both a ``yes''
and a ``no'' answer yield the same acceptance or rejection behavior),
then we lose at least one mind change on every path 
in $H_M(x)$. 
\end{fact}

So from now on, in light of Fact~\protect\ref{f:2}, let $M^{A[1]:B[1]}(x)$ 
be a $\pjk$ machine that has, on input $x$, one of the following four 
acceptance schemes (the scheme may depend on the input).

\begin{description}
\item[(1)] $M$ accepts if and only if exactly one of the two sequential 
queries is answered ``yes.''
\item[(2)] $M$ accepts if and only if either both or neither of the two 
asked queries is answered ``yes.''
\item[(3)] $M$ accepts if and only if the second query is answered 
``yes.''
\item[(4)] $M$ accepts if and only if the second query is answered 
``no.''
\end{description}
 
\begin{fact}\label{f:1}
If $p$ is an ascending path in $H_M(x)$ such that $p$ contains adjacent 
vertices $v$ and $v+u_d$ such that 
$$d\le j \makebox{ and the } (d')\makebox{th component of } v \makebox{ is 
0 for some } d'<d,$$
then $p$ loses a mind change.
\end{fact}

\noindent {\bf Proof}\quad
Since $A \in \bh_j$ and thus $A=A_1-(A_2-(\cdots-(A_{j-1}-A_j)\cdots ))$ 
and there 
is a 0 in the $(d')$th component of $v$ and $v+u_d$, both vertices yield the 
same answer to ``$q_1(x) \in A$?'' The 1 in the $d$th component of $v+u_d$ has 
no effect at all on the answer to ``$q_1(x) \in A$?''~and so on the 
outcome of $M^{A[1]:B[1]}(x)$. 
Hence, both vertices have the same label and $p$ loses a mind change.~\qed

Similarly, one can prove that if $p$ is an ascending path and $p$ 
contains two adjacent vertices $v$ and $v+u_d$ such that 
$j<d'<d \le j+k$ and the $(d')$th component of $v$ is 0 or 
$j+k<d'<d \le j+2k$ and the $(d')$th component of $v$ is 0 
then $p$ also loses one mind change. 

Furthermore, in light of Fact~\protect\ref{f:1}, 
let us focus only on paths $p$ that 
change their first $j$, second $k$, and last $k$ dimensions from the
smallest to the 
highest dimension in each group.
This allows us to simplify the description of paths as follows. Let 
$e_1$ be the following operator on $H$:
\begin{quote}
$e_1((a_1,\cdots,a_{j+2k}))=$ \\\nopagebreak
$\mbox{~~~~~~}
\left\{
\begin{array}{l}
\protect (a_1,\cdots,a_{i-1},1,\cdots,a_{j+2k}) \mbox{~~if $i\le j$, 
$a_i=0 \land (\forall j:j<i)[a_j=1$]}\\
\protect (a_1,\cdots,a_{j+2k}) \mbox{~~~~~~~~~~~~~~~~~otherwise.}
\end{array}
\right. $
\end{quote}

The operators $e_2$ and $e_3$ act on the index groups $(j+1,\cdots,j+k)$ and 
$(j+k+1,\cdots,j+2k)$, respectively, in the same manner: the zero 
component with smallest index among the zero components is incremented by 1.
The only reasonable paths to consider are those emerging from 
repeated applications of $e_1$, $e_2$ and $e_3$ to $(0,\cdots,0)$.
We will use $(e_{i_1},e_{i_2}, \cdots ,e_{i_{j+2k}})$ to denote
the path with vertices 
$v_0=(0,\cdots,0)$, $v_1=e_{i_1}(v_0), v_2=e_{i_2}(v_1),\cdots$ ,
\mbox{$v_{j+2k}=e_{i_{j+2k}}(v_{j+2k-1})=(1,1, \cdots ,1)$}.

\noindent The next fact gives 
sufficient conditions for an ascending path to lose 
a mind change, namely:

\begin{fact}\label{f:3}
On any ascending path $p$ a mind change loss occurs if:
\begin{description}
\item[Case 1.1] there is an $e_2$ after an odd number of $e_1$'s in the 
description of $p$, or
\item[Case 1.2]there is an $e_3$ after an even number of $e_1$'s in the 
description of $p$, or
\item[Case 2] the description of $p$ contains a sequence of odd length 
at least 3 that starts and ends with $e_1$ and contains no other $e_1$'s.
\end{description}
\end{fact}

\noindent {\bf Proof} \quad
We will call the occurrence of Case 1.1 (Case 1.2) in $p$ an ``$e_2$-loss'' 
(``$e_3$-loss'') and the occurrence of Case 2 an ``odd episode.'' In general 
we call a subpath of $p$ of length at least 3 that starts and ends 
with $e_1$ and contains no other $e_1$ an episode.

Intuitively $p$ loses a mind change in the case of Case~1.1 (1.2), since in 
the actual computation $M(x)$ does not really ask query $q_2(x)$ ($q_3(x)$) 
and so a change in the answers to the $k$ underlying $\np$ queries 
of $q_2(x)$ ($q_3(x)$) does not affect the outcome of the overall computation. 

Intuitively in Case~2 the following argument holds. If the description 
of $p$ contains an odd episode, say starting with 
$e_{i_l}=e_1$ and ending with $e_{i_{l'}}=e_1$, then 
$v_{l-1},v_l,\cdots,v_{l'}$ form an even-length subpath $p'$ of $p$. 
If the odd episode contains both $e_2$'s and $e_3$'s then note that Case~1 
applies and we are done. In fact due to Case~1, we may hence forward 
assume the odd episode, between the starting and the ending $e_1$'s, 
has only $e_2$'s (respectively $e_3$'s), if we have an even (respectively odd)
 number of $e_1$'s up to and including the $e_1$ starting the odd episode. 
 So in this case $v_{l-1}$ and $v_{l'}$ have the same label Accept/Reject. 
The acceptance behavior 
of $M^{A[1]:B[1]}(x)$ due to $v_{l-1}$ and $v_{l'}$ is the same, because 
after 
two $e_1$'s the answer to ``$q_1(x) \in A$?'' is the same as it was 
before the two $e_1$'s, and the $e_2$'s ($e_3$'s) have not influenced 
the answer to $q_3(x)$ ($q_2(x)$). 
Thus we have a subpath of even length, namely $v_{l-1},v_l,\cdots,v_{l'}$, 
whose starting point and endpoint 
have the same Accept/Reject label. To assign to each vertex of this path 
an Accept/Reject label in such a way that no mind changes are lost is 
equivalent to the impossible task of 2-coloring an odd cycle. 
Hence we lose at least one mind change for every occurrence of an 
odd episode.~\qed

Before proving the main theorem of this section, we show the 
following lemma, Lemma~\ref{l:mind}, which
tells how many mind changes $\pjk$ can handle.
We say a complexity class $\pjk$ can handle exactly $m$ 
mind changes if and only if (a)~no $\pjk$ machine has 
more than $m$ mind changes and (b)~there is a specific 
$\pjk$ machine that has $m$ mind changes. 
It is known (see, e.g., 
\cite{cai-gun-har-hem-sew-wag-wec:j:bh1,koe-sch-wag:j:diff,bei:j:bounded-queries}) 
that 
$\redttnp{k}$ can handle exactly $k$ mind changes.

\begin{lemma} \label{l:mind}
The class $\pjk$ can handle exactly $m$ mind changes, where
$$
m = \left\{
\begin{array}{l}
\protect\ j+2k-1 \mbox{~~~~~~~~~~~~~~~~~~~~~ if $j$ is even and $k$ is odd}\\
\protect\ j+2k \mbox{~~~~~~~~~~~~~~~~~~~~~~~~~~~otherwise.}
\end{array}
\right.$$
\end{lemma}

\noindent{\bf Proof} \quad
We first consider the case in which $j$ is even and $k$ is odd. 

We want to argue that for every $\pjk$ machine $M$ and every 
$x \in \Sigma^*$,
on every ascending path in the 
\makebox{$j+2k$} dimensional, appropriately labeled, hypercube $H_M(x)$ there 
are at most $j+2k-1$ mind changes.
Let $x\in \Sigma^*$ and $M$ be a $\pjk$ machine with the oracles $A$ and $B$. 
Due to Facts~\ref{f:1} and \ref{f:2}, it suffices to consider 
a $\pjk$ machine 
$M$ with one of the four previously mentioned acceptance schemes on input $x$ 
and to show that every path $p$ having the introduced description loses 
at least one mind change.
Let $M(x)$ be such a machine and $p$ be such a path. There are two 
possibilities.

\begin{description}

\item[Case A] The description of $p$ contains an $e_2$-loss or an $e_3$-loss.\\
According to Fact~\ref{f:3}, $p$ loses at least one mind change.

\item[Case B] The description of $p$ contains neither an $e_2$- loss nor an 
$e_3$-loss.\\
Hence the description of $p$ consists of blocks of consecutive $e_2$'s 
and $e_3$'s separated by blocks of $e_1$'s.
Since the description of $p$ contains $k$ $e_3$'s and $k$ is odd, there 
is a block of $e_3$'s of odd size in $p$. Since we have no $e_3$-loss and 
$j$ is even this block is surrounded by $e_1$'s. Thus we have an odd 
episode in the description of $p$ and, according to Fact~\ref{f:3}, $p$ loses a 
mind change.
\end{description}

So no $\pjk$ machine can realize more than $j+2k-1$ mind changes.  

It remains to show that there is a deterministic $\pjk$ machine and an
input $x \in \Sigma^*$ such 
that in the associated hypercube $H_M(x)$ there is a path having exactly 
$j+2k-1$ mind changes.

Let us consider the path $p_0$, 
$$p_0=(\underbrace{e_2,e_2,\cdots,e_2}_k,\underbrace{e_1,e_1,\cdots,e_1}_{j-1}
,\underbrace{e_3,e_3,\cdots,e_3}_k,e_1).$$

Consider the deterministic oracle machine $W$ that 
asks two sequential queries and accepts an input $x$ if and only if 
the second query of $W(x)$ was answered ``yes'' (acceptance scheme (3)).
We know as just shown that all ascending paths of $H_W(x)$ have at most $j+2k-1$
mind changes.
Note that for every $x \in \Sigma^*$ the path $p_0$ loses only one 
mind change and thus $\pjk$ can 
handle exactly $j+2k-1$ mind changes.

This completes the proof of the case ``$j$ is even and $k$ is odd.''
We now turn to the ``$j$ is odd or $k$ is even'' case of the lemma being
proven.

Since our hypercube has (in all cases) $j+2k$ dimensions,
certainly $\pjk$ can handle (in all cases) no more than $j+2k$ mind changes.

If $j$ is odd, we consider the path 
$$p_1=(\underbrace{e_2,e_2,\cdots,e_2}_k,\underbrace{e_1,e_1,\cdots,e_1}_j
,\underbrace{e_3,e_3,\cdots,e_3}_k)$$ 
and--- 
using the acceptance 
scheme numbering introduced just after
Fact~\ref{f:1}---we consider 
the machine having for every input $x$ acceptance scheme~(3) or~(1) 
for $k$ odd or even, respectively.
If $j$ is even and $k$ is even, we consider path $p_0$ and 
we consider the machine 
having acceptance scheme~(1) for every input.

In each of these cases the considered machine changes its mind along the 
associated path exactly $j+2k$ times. 
Hence for $j$ odd or $k$ even the class $\pjk$ 
can handle exactly $j+2k$ mind changes.~\qed

Now we are ready to prove our main theorem of this section.
\begin{theorem} \label{t:main}
For $j,k \geq 1$,
$$
\pjk = \left\{
\begin{array}{l}
\protect\redttnp{j+2k-1} \mbox{~~~~~~~~~~~~~~~~~~~~~if $j$ is even and 
$k$ is odd}\\
\protect\redttnp{j+2k}    \mbox{~~~~~~~~~~~~~~~~~~~~~~~~otherwise.}
\end{array}
\right.$$
\end{theorem}

\noindent{\bf Proof} \quad 
In order to avoid unnecessary case distinctions we prove the fact for 
arbitrary $j$ and $k$ and simply denote the appropriate number of mind 
changes by $m$, namely (see Lemma \ref{l:mind})
$j+2k-1$ if $j$ is even and $k$ is odd and $j+2k$ otherwise.
First we would like to show that $\pjk \subseteq \redttnp{m}$. 
We show this by explicitly giving the appropriate truth-table reduction.

Let $A \in \pjk$ and let $m$ be the number of mind changes (according to
Lemma \ref{l:mind}) the class $\pjk$ can handle. 
Let $M$ be a deterministic oracle machine, witnessing $A \in \pjk$, via the
sets $S_1 \in \bh_j$ and $S_2 \in \bh_k$.
As noted by Beigel~\cite{bei:j:bounded-queries}, the set \makebox{$Q=
\{\langle x,k \rangle \condition M(x)$ has at least $k$ mind changes$\}$} 
is an $\np$ set. Note that if $M^{S_1[1]:S_2[1]}(x)$ on a particular 
input $x$ rejects (respectively accepts) if both queries have the 
answer ``no'' then $M^{S_1[1]:S_2[1]}(x)$ accepts if and only if 
the node (of the implicit hypercube) associated with the actual 
answers has an odd (respectively even) number of mind changes.

Define the variables $o,y_1,y_2,\cdots,y_m$ and the $m$-ary boolean 
function $\alpha$:\\
\mbox{~~~~~~~~~~~~~~~$o=0$  if $M^{S_1[1]:S_2[1]}(x)$ rejects if both queries 
are answered ``no,''}\\
\mbox{~~~~~~~~~~~~~~~$o=1$ if $M^{S_1[1]:S_2[1]}(x)$ accepts if both queries 
are answered ``no,''}\\ 
\mbox{~~~~~~~~~~~~~~~~~~~~~~~~~~~~~~~~~~~$y_1=\langle x,1 \rangle$,}\\
\mbox{~~~~~~~~~~~~~~~~~~~~~~~~~~~~~~~~~~~$y_2=\langle x,2 \rangle$,}\\
\mbox{~~~~~~~~~~~~~~~~~~~~~~~~~~~~~~~~~~~$y_3=\langle x,3 \rangle$,}\\
\mbox{~~~~~~~~~~~~~~~~~~~~~~~~~~~~~~~~~~~~~~~~~~~~$\vdots$}\\
\mbox{~~~~~~~~~~~~~~~~~~~~~~~~~~~~~~~~~~~$y_m=\langle x,m \rangle$,}\\
and
\mbox{~~~~~~~~~~~$\alpha(z_1,z_2,\cdots,z_m)=1 \iff 
(\max\{l \condition z_l=1\}+o) \equiv 1 \pmod{2}$.}

Clearly we can compute the just defined variables for a given $x$ and 
also evaluate the function $\alpha$ at \mbox{$(\chi_Q(y_1),\chi_Q(y_2),
\cdots,\chi_Q(y_m))$} in polynomial time.
And finally we have \makebox{$x \in A \iff \alpha(\chi_Q(y_1),\chi_Q(y_2),
\cdots,\chi_Q(y_m))=1$}. Thus $A \in \redttnp{m}$.

It remains to show that \mbox{$\redttnp{m} \subseteq \pjk $}. Recall 
\mbox{$\redttnp{k}=\redttbhk{1}$} from Lemma~\ref{l:tt}. 
Since the class 
$\pjk $ is closed under $\leq^p_{{\scriptsize\mbox{1-tt}}}$ reductions 
it suffices to prove $\bh_m \subseteq \pjk$.

So let $B \in \bh_m$. Following Cai et al.~\cite{cai-gun-har-hem-sew-wag-wec:j:bh1} we may assume that 
the set $B$ is of the form
\makebox{$B=B_1-(B_2-(B_3-( \cdots -(B_{m-1}-B_m)\cdots)))$} with 
\mbox{$B_1,B_2,\cdots,B_m \in \np$} and \mbox{$B_m \subseteq 
\cdots \subseteq B_2 \subseteq B_1$}.

We show $B \in \pjk$ by using ideas of the second part of the proof of 
Lemma 4.1, namely by implementing the specific good path $p_0$, 
respectively $p_1$.
$B$ is accepted by a $\pjk$ machine $M$ as follows:

\begin{description}
\item[Case 1] $j$ is odd.\\
Define the two oracle sets $O_1$ and $O_2$:\\
\makebox{$O_1=B_{k+1}-(B_{k+2}-(\cdots -(B_{k+j-1}-B_{k+j})\cdots ))$, and}\\
\makebox{$O_2=\{\langle y,2 \rangle \condition y \in 
B_1-(B_2-( \cdots -(B_{k-1}-B_k)\cdots ))\}$}\\
\makebox{~~~~~~~~~~$\cup \{\langle y,3 \rangle 
\condition y \in B_{j+k+1}-(B_{j+k+2}-(\cdots -(B_{j+2k-1}-B_{j+2k})\cdots))\}$.}\\

Note that $O_1 \in \bh_j$ and $O_2 \in \bh_k$.
On input $x$ $M$ first queries ``$x \in O_1$.'' In case of 
a ``no'' answer $M(x)$ queries $\langle x,2 \rangle \in O_2$ and 
in case of a ``yes''answer to the first query $M(x)$ asks 
$\langle x,3 \rangle \in O_2$.

\begin{description}
\item[Case 1.1] $k$ is odd.\\
$M(x)$ accepts if and only if the second query is answered ``yes.''
\item[Case 1.2] $k$ is even.\\
$M(x)$ accepts if and only if exactly one of the two queries is 
answered ``yes.''
\end{description}

\item[Case 2] $j$ is even.\\
Define the two oracle sets $O_1$ and $O_2$:\\
\makebox{$O_1=B_{k+1}-(B_{k+2}-(\cdots -(B_{k+j-1}-B_m)\cdots ))$, and}\\
\makebox{$O_2=\{\langle y,2 \rangle \condition y \in 
B_1-(B_2-( \cdots -(B_{k-1}-B_k)\cdots ))\}$}\\
\makebox{~~~~~~~~~~$ \cup \{\langle y,3 \rangle 
\condition y \in B_{j+k}-(B_{j+k+1}-(\cdots -(B_{m-2}-B_{m-1})\cdots))\}$.}\\

Note that $O_1 \in \bh_j$ and $O_2 \in \bh_k$.
On input $x$ $M$ first queries ``$x \in O_1$.'' In case of a ``no'' 
answer $M(x)$ queries $\langle x,2 \rangle \in O_2$ and in case of a 
``yes'' answer to the first query $M(x)$ asks $\langle x,3 \rangle \in O_2$.
\begin{description}

\item[Case 2.1] $k$ is odd.\\
$M(x)$ accepts if and only if the second query is answered ``yes.''
\item[Case 2.2] $k$ is even.\\
$M(x)$ accepts if and only if exactly one of the two queries is 
answered ``yes.''~\qed
\end{description}
\end{description}

It is interesting to note which properties of NP are actually required
in the above proof for the result to hold. The proof essentially rests
on the fact that the key set $Q$ (describing that, for given $x$ and
$m$, the $\pjk$ machine $M$ on input $x$ has at least $m$ mind
changes) is an NP set. So considering an arbitrary underlying class
${\cal C}$, for proving $Q \in {\cal C}$ it suffices to note that $Q$
is in the class $\exists^{b} \cdot 
{\rm R}_{\rm c\mbox{-}btt}^{p}({\cal C})$,\footnote{%
\protect\singlespacing
Here, $\leq_{\rm c\mbox{-}btt}^{p}$ denotes the conjunctive bounded truth-table
reducibility, and for any class ${\cal K}$, 
$\exists^{b} \cdot {\cal K}$ is defined to be the
class of languages $A$ for which there exists a set $B \in {\cal K}$ 
and a constant bound $m$ such that $x \in A$ if and only if there exists 
a string $y$ of length at most $m$ with $\langle x, y \rangle \in B$.
} 
and to assume that ${\cal C}$ be closed under $\exists^{b}$ and
conjunctive bounded truth-table reductions. Indeed, the $\exists^{b}$
quantifier describes that there is a path in the boolean hypercube
$H_{M}(x)$, and via the $\leq_{\rm c\mbox{-}btt}^{p}$-reduction it can be
checked that this path is an ascending path and all the answers the
vertices on that path claim to be ``yes'' answers indeed correspond to
query strings that belong to the class ${\cal C}$. Similar
observations have been stated in earlier papers
\cite{bei:j:bounded-queries,bei-cha-ogi:j:difference-hierarchies}.
In terms of the present paper, note in particular
that the assertion of Theorem~\ref{t:main} holds true
for all classes ${\cal C}$ closed under union,
intersection, and polynomial-time many-one reductions.  
$\mbox{C}_{=}\mbox{P}$, R, and FewP all have these 
closure properties, to name just a few examples. 
If the underlying class ${\cal
C}$ is closed under polynomially bounded $\exists$
quantification and unbounded conjunctive truth-table reductions, it is
not hard to see that this analysis can even be done safely up to the case
of logarithmically bounded query classes, as the number of paths in
the hypercube is polynomial and thus generates a polynomial-sized
disjunction.

Theorem~\ref{t:main} allows us to derive a relationship between classes of the form
$\pjk$ and $\p^{{\rm BH}_j [1]:{\rm BH}_k [1]_+}$. Classes of the latter form were 
studied in~\cite{agr-bei-thi:tOutByFSTTCSConf:modulo-information}.

\begin{corollary}\label{c:rel-r-h}
For every $j,k \geq 1$,
$$
{\rm R}^p_{1{\scriptsize\mbox{-tt}}}(\p^{{\rm BH}_j [1]:{\rm BH}_k [1]_+}) = 
\left\{
\begin{array}{l}
\protect{\p^{{\rm BH}_{j-1} [1]:{\rm BH}_k [1]}} 
\mbox{~~~~~~~~~~~~if $j$ is odd and $k$ is even}\\
\protect\pjk \mbox{~~~~~~~~~~~~~~otherwise.}
\end{array}
\right.$$
\end{corollary}
The proof is immediate by 
the results of Theorem~\ref{t:main} of this paper and 
Theorem~5.7 and Lemma~5.9 of \cite{agr-bei-thi:tOutByFSTTCSConf:modulo-information}.

From Theorem~\ref{t:main} we can immediately conclude that order
matters for queries to the boolean hierarchy unless the boolean
hierarchy itself collapses.  
\begin{corollary}
\begin{enumerate}
\item If $ (j=k) \lor (j$ is even and $k = j+1)$, $1\leq j \leq k$, then
$\pjk = \pkj$.
\item Unless the boolean hierarchy (and thus the polynomial
hierarchy) collapses: for every $1\leq j \leq k$, 
$\pjk \neq \pkj$
unless 
$ (j=k) \lor (j$ is even and $k = j+1)$.
\end{enumerate}
\end{corollary}

The corollary holds, in light of the theorem, simply because the boolean
hierarchy and the truth-table hierarchy are
interleaved~\cite{koe-sch-wag:j:diff} in such a way that the boolean
hierarchy levels are sandwiched between levels of the
bounded-truth-table hierarchy, and thus if two different levels of the
bounded-truth-table hierarchy are the same (say levels
$r$ and $s$, $r<s$), then some level (in particular, $\bh_{r+1}$) of the
boolean hierarchy is closed under complementation, and thus, by the
downward separation property of the boolean
hierarchy~\cite{cai-gun-har-hem-sew-wag-wec:j:bh1}, the boolean
hierarchy would collapse.  Furthermore,
Kadin~\cite{kad:joutdatedbychangkadin:bh} has shown that if the
boolean hierarchy collapses then the polynomial hierarchy collapses,
and Wagner, and Chang and Kadin, and Beigel, Chang, and Ogihara have
improved the strength of this
connection~\cite{wag:t:n-o-q-87version,wag:t:n-o-q-89version,cha-kad:j:closer,bei-cha-ogi:j:difference-hierarchies}.
The strongest known connection is: If $\bh_q = \cobh_q$, then $\ph =
{\left( \p_{(q-1)\mbox{\scriptsize{}-tt}}^{\rm NP}\right)}^{\rm
NP}$~\cite{bei-cha-ogi:j:difference-hierarchies}, where ${\left(
\p_{m\mbox{\scriptsize{}-tt}}^{\rm NP}\right)}^{\rm NP}$ denotes the
class of languages accepted by P machines given $m$-truth-table access
to an $\npnp$ oracle and also given unlimited access to an NP oracle
(note that 
${\left( \p_{1\mbox{\scriptsize{}-tt}}^{\rm NP}\right)}^{\rm NP}$ is 
equal to $\p^{ {\rm NP^{NP}}[1]: {\rm NP}}$
as leading NP
queries can be absorbed into the $\npnp$ query).  

In light of this discussion, we can make more clear exactly what collapse 
is spoken of in the second part of the above corollary.  In particular, 
the collapse of the polynomial hierarchy is (at least) to 
${\left( \p_{(k+2j)\mbox{\scriptsize{}-tt}}^{\rm NP}\right)}^{\rm
NP}$.\footnote{\protect\singlespacing
Though one level is gained by the $q-1$ in the 
\protect\cite{bei-cha-ogi:j:difference-hierarchies} connection between the boolean hierarchy
and the polynomial hierarchy, one level is lost in the collapse of
the boolean hierarchy that follows from a given collapse in the 
truth-table hierarchy.  We speculate that it might be 
possible for the $k+2j$ claim to be strengthened
by one level by applying the 
\protect\cite{bei-cha-ogi:j:difference-hierarchies} technique directly to the 
truth-table hierarchy.}

Of course, Theorem~\ref{t:main} applies far more generally.  From it,
for any $j$, $k$, $j'$, and $k'$, one can either immediately conclude
equality, or can immediately conclude that the classes are not equal
unless the polynomial hierarchy collapses to ${\left(
\p_{(\min(\alpha(j,k),\,\alpha(j',k')))\mbox{\scriptsize{}-tt}}^{\rm
NP}\right)}^{\rm NP}$, where $\alpha(a,b)$ equals $a+2b-1$ if $a$ is
even and $b$ is odd and $a+2b$ otherwise.

The point of Theorem~\ref{t:main} is that from
the even/odd structure of $\pjk$ classes one can immediately
tell their number of mind changes, and thus their strength, without 
having to do a separate, detailed, mind change analysis
for each $j$ and $k$ pair.  However, note that one can, via a time-consuming
but mechanical procedure, analyze almost any class with a query 
tree structure (namely by looking at the full tree of 
possible queries and answers, and for each of the huge number of 
possible ways its leaves can each be labeled accept-reject 
compute the number of mind changes that labeling creates,
and then look at the maximum over all these numbers).  
For example, one can quickly see that one query to
DP followed by 4-tt access to NP yields exactly the languages
in $\redttnp{10}$.

\section{General Case} \label{section:general}

In the previous section, we studied classes of the form $\pjk$. We 
completely characterized them in terms of reducibility hulls of $\np$ and 
noted that in this setting the order of access to different oracles matters 
quite a bit.
What can be said about, for example, the class 
${\rm P}^{ {\rm BH}_j [1] : {\rm BH}_k [1] : {\rm BH}_l [1]}$?
Is it equal to
$\pjkl$?  (We'll see that the answer is ``no'' in certain cases.) 
Even more generally, what can
be said about the classes of languages that are accepted by 
deterministic oracle machines with tree-like query structures and 
with each query 
being made to a (potentially) different oracle
from a (potentially) different level of the boolean hierarchy?
Is it possible that with a more complicated query structure we might lose 
even more than the one mind change lost in the case of 
${\rm P}^{ {\rm BH}_j [1] : {\rm BH}_k [1]}$ with $j$ even and $k$ odd?
(From the results of the section, it will be clear 
that the answer to this question
is ``yes'';  mind changes can, in certain specific 
circumstances, accumulate.)

First of all, 
we can immediately derive a characterization of the class $\pjkl$ from 
the results of the previous section, namely:

\begin{theorem}\label{t:jkl}
For $j,k,l \geq 1$,
$$
\pjkl = \left\{
\begin{array}{l}
\protect\redttnp{j+k+l-1} \mbox{~~~~~~~~~~~~~~~~~~~~~if $j$ is even and 
$l$ is odd}\\
\protect\redttnp{j+k+l}    \mbox{~~~~~~~~~~~~~~~~~~~~~~~~otherwise.}
\end{array}
\right.$$
\end{theorem}

\noindent {\bf Proof} \quad
Note that in Lemma~\ref{l:mind} we handle the special case of $k=l$.
However, notice that the mind change loss for 
$j$ even and $k$ odd is due only to the fact that the query made after the first 
query is answered ``yes'' 
is made to an oracle from an odd level, namely $k$, 
of the boolean hierarchy. 
In particular the mind change loss is not tied to the query we ask in case the 
first query is answered ``no.''
Thus we have 
\begin{description}
\item[Claim] The class $\pjkl$ can handle exactly $m$ mind changes where
$$
m = \left\{
\begin{array}{l}
\protect\ j+k+l-1 \mbox{~~~~~~~~~~~~~~~~~~~~~ if $j$ is even and $l$ is odd}\\
\protect\ j+k+l \mbox{~~~~~~~~~~~~~~~~~~~~~~~~~~~otherwise.}
\end{array}
\right.$$
\end{description}

\noindent
Similarly to the proof of 
Theorem~\ref{t:main} one can now show the equality we claim.~\qed

Note that for every $j,k,l \geq 1$, we obviously have 
$$\pjkl=\pttjttkttl$$ 
and thus the following corollary holds. 

\begin{corollary}\label{c:jkl}
For $j,k,l \geq 1$,
$$
\pttjttkttl = \left\{
\begin{array}{l}
\protect\redttnp{j+k+l-1} \mbox{~if $j$ is even and 
$l$ is odd}\\
\protect\redttnp{j+k+l}    \mbox{~~~~otherwise.}
\end{array}
\right.$$
\end{corollary}

The last corollary is the key tool 
to use in evaluating any class of languages that 
are accepted by deterministic 
oracle machines with tree-like query structures and 
with each query being made to a (potentially) different oracle
from a (potentially) different level of the boolean hierarchy.

We formalize some notions to use in studying this.  
Let $T$ be a binary tree, not necessarily complete, such that each 
internal node $v_i$ (a)~has exactly two children, and (b)~is labeled 
by a natural number $n_i$ (whose purpose will be explained below).  
For such a tree $T$, define $f_T$ by $f_T(v_i)=n_i$.  Henceforward,
we will write $f$ for $f_T$ in contexts in which $T$ is clear.
Let $root_T$ 
be the root of the tree 
(we will assign to this node the name $v_1$)
and let $LT_T$ and $RT_T$ respectively be the left and  
right subtrees of the root. 
We will denote the class of sets that are accepted by a deterministic 
oracle machine with a $T$-like query structure by ${\rm P}^{(T)}$. 
Here the structure of the tree $T$ gives the potential computation 
tree of every ${\rm P}^{(T)}$ machine in the sense that inductively 
if a query at node $v$ is answered ``no'' (``yes'') we keep on moving through 
the tree in the left (right) subtree of $v$. And at each internal node $v_i$ 
of $T$ the natural number $n_i$ gives the level of the 
boolean hierarchy from which the oracle queried at that node
is taken.

\begin{figure}[tp]
\begin{center}
  \leavevmode
  \unitlength3mm
  \begin{picture}(36,32)
    \put(0,16){\circle{2.0}\makebox(0,0){$v_1$}}
    \put(0,14){\makebox(0,0){$n_1=2$}}
    \put(12,24){\circle{2.0}\makebox(0,0){$v_3$}}
    \put(12,22){\makebox(0,0){$n_3=4$}}    
    \put(12,8){\circle{2.0}\makebox(0,0){$v_2$}}
    \put(12,6){\makebox(0,0){$n_2=2$}}
    \put(24,28){\circle{2.0}\makebox(0,0){$v_5$}}
    \put(24,26){\makebox(0,0){$n_5=3$}}
    \put(24,20){\circle{2.0}\makebox(0,0){$v_4$}}
    \put(24,18){\makebox(0,0){$n_4=1$}}
    \put(24,12){\makebox(0,0){leaf}}
    \put(24,4){\makebox(0,0){leaf}}
    \put(36,30){\makebox(0,0){leaf}}
    \put(36,26){\makebox(0,0){leaf}}
    \put(36,22){\makebox(0,0){leaf}}
    \put(36,18){\makebox(0,0){leaf}}

    \put(0.8,16.45){\line(3,2){10.4}}
    \put(0.8,15,55){\line(3,-2){10.4}}
    \put(12.8,24.45){\line(3,1){10.2}}
    \put(12.8,23.55){\line(3,-1){10.2}}
    \put(12.8,8.45){\line(3,1){9}}
    \put(12.8,7.55){\line(3,-1){9}}
    \put(24.8,28.45){\line(6,1){9}}
    \put(24.8,27.55){\line(6,-1){9}}
    \put(24.8,20.45){\line(6,1){9}}
    \put(24.8,19.55){\line(6,-1){9}}

  \end{picture}
\end{center}
\caption{Tree $\cal T$}
\label{fig:ctt-tree-3}

\end{figure}

For example consider the tree $\cal T$ (see Figure~\ref{fig:ctt-tree-3}), 
in which $f(v_1)=2$, $f(v_2)=2$, $f(v_3)=4$, $f(v_4)=1$, and $f(v_5)=3$.
A ${\rm P}^{({\cal T})}$ machine works as follows. The first query is made to a
$\rm DP$ oracle. If the answer to that first query
is ``no'' a second query is made to the 
$\rm DP$ oracle associated with $v_2$, 
and if the answer to the first query is  ``yes'' the 
second query is made to the $\bh_4$ oracle associated with $v_3$.
A third query is made only if the answer to the first 
query is ``yes''; in this case, 
the oracle set of the third query is in $\np$
if the answer to the second query is ``no,'' and is in $\bh_3$ if the 
answer to the second query is ``yes.''
Note that for every input $x \in \Sigma^*$ every ${\rm P}^{({\cal T})}$ 
machine 
$M(x)$ assigns a label A (Accept) or 
R (Reject) to each leaf of $\cal T$ with its own specific acceptance 
behavior (which,
in particular, may depend on $x$).

If $T$ is the complete tree of depth 1 (i.e., a root plus 
two leaves), then by definition $m(T)=f(root_T)$ , and otherwise define
$$
m(T)= \left\{
\begin{array}{l}
\protect f(root_T)+m(LT_T)+m(RT_T)-1 \mbox{~~~~if $f(root_T) \equiv 0 \pmod{2}$ and}\\
\protect \mbox{~~~~~~~~~~~~~~~~~~~~~~~~~~~~~~~~~~~~~~~~~~~~~~~~~~~~~$m(RT_T) \equiv 1 
\pmod{2}$}\\
\protect f(root_T)+m(LT_T)+m(RT_T) \mbox{~~~~~~~~~otherwise.}
\end{array}
\right.$$ 

For our example tree
$\cal T$ we have $m({\cal T})=10$. 
The main theorem of this section will prove 
$m(T)$ determines the number of bounded truth-table accesses
to NP that completely characterize the
class $\p^{(T)}$.  It follows from the main
theorem that, for example, ${\rm P}^{({\cal T})}=\redttnp{10}$.

\begin{theorem} \label{t:maingeneral}
$ {\rm P}^{(T)}=\redttnp{m(T)}$.
\end{theorem}

\noindent {\bf Proof} \quad
The proof consists of an obvious induction over the depth $d$ of the tree.
Note that the correctness of the base case of the induction, $d=2$, 
is given by 
Theorem~\ref{t:jkl}.  
The proof of the inductive step follows immediately
from the obvious fact that 
$${\rm P}^{(T)}={\rm P}^{{\rm BH}_{f(root_T)}:{\rm P}^{(LT_T)},{\rm P}^{(RT_T)}},$$
combined with 
Lemma~\ref{l:tt} ($\redttnp{k}=\redttbhk{1}$)
and Corollary~\ref{c:jkl}.~\qed

Finally, we mention that a study of
query order in the polynomial hierarchy
(as opposed to the boolean hierarchy) has
very recently been
initiated by E.~Hemaspaandra, L.~Hemaspaandra, and
H.~Hempel~(\cite{hem-hem-hem:ctoappear:query-order-ph},
see 
also~\cite{wag:t:parallel-difference,bei-cha:c:commutative-queries})
and this study has led to a somewhat surprising downward translation
result:  For $k>2$, 
$\sigmak = \pik \iff
\p^{\sigmak[1]} = 
\p^{\sigmak[2]}$~(\cite{hem-hem-hem:jtoappear:downward-translation},
see also the extensions obtained 
in~\cite{buh-for:t:two-queries,hem-hem-hem:t:translating-downwards}).
Query order 
(see also the survey~\cite{hem-hem-hem:jtoappear:query-order-survey})
has also recently proven useful in 
studying the structure of complete 
sets~\cite{hem-hem-hem:jtoappear:sn-1tt-np-completeness}
and in characterizing bottleneck-computation
classes~\cite{her:c:transformation-monoid-acceptance}.

{\samepage
\begin{center}
{\bf Acknowledgments}
\end{center}
\nopagebreak
\indent
We thank Edith Hemaspaandra, Johannes K\"obler, and J\"org Vogel for
helpful conversations.  We thank Johannes K\"obler for providing an
advance copy of~\cite{gre-koe-reg-sch-tor:j:middle-bit}.  We are
extremely indebted to J\"org Rothe for his generous help.  
The insights of
the paragraph after the proof of Theorem~\ref{t:main} are
due to him, and appear here with his kind permission.  He also
made countless invaluable suggestions throughout this project, and proofread
an earlier version of this paper.
We also are deeply grateful to editor Ker-I Ko 
and two anonymous referees for their invaluable suggestions 
regarding the organization of the paper and
a very nice proof simplification for Section~\ref{section:general}.
}%


\begin{thebibliography}{10}

\bibitem{agr-bei-thi:tOutByFSTTCSConf:modulo-information}
{\sc M.~Agrawal, R.~Beigel, and T.~Thierauf}, {\em Modulo information from
  nonadaptive queries to {NP}}, Tech. Report 96-001, Electronic Colloquium on
  Computational Complexity, Jan. 1996.

\bibitem{bal-dia-gab:b:sctI-2nd-ed}
{\sc J.~Balc\'{a}zar, J.~D\'{\i}az, and J.~Gabarr\'{o}}, {\em Structural
  Complexity I}, EATCS Texts in Theoretical Computer Science, Springer-Verlag,
  2nd~ed., 1995.

\bibitem{bei:j:bounded-queries}
{\sc R.~Beigel}, {\em Bounded queries to {SAT} and the boolean hierarchy},
  Theoretical Computer Science, 84 (1991), pp.~199--223.

\bibitem{bei-cha:c:commutative-queries}
{\sc R.~Beigel and R.~Chang}, {\em Commutative queries}, in Proceedings of the
  5th Israeli Symposium on Theory of Computing and Systems, IEEE Computer
  Society Press, June 1997, pp.~159--165.

\bibitem{bei-cha-ogi:j:difference-hierarchies}
{\sc R.~Beigel, R.~Chang, and M.~Ogiwara}, {\em A relationship between
  difference hierarchies and relativized polynomial hierarchies}, Mathematical
  Systems Theory, 26 (1993), pp.~293--310.

\bibitem{bov-cre:b:complexity}
{\sc D.~Bovet and P.~Crescenzi}, {\em Introduction to the Theory of
  Complexity}, Prentice Hall, 1993.

\bibitem{bru-jos-you:j:strong}
{\sc D.~Bruschi, D.~Joseph, and P.~Young}, {\em Strong separations for the
  boolean hierarchy over {RP}}, International Journal of Foundations of
  Computer Science, 1 (1990), pp.~201--218.

\bibitem{buh-for:t:two-queries}
{\sc H.~Buhrman and L.~Fortnow}, {\em Two queries}, Tech. Report 96-20,
  University of Chicago, Department of Computer Science, Chicago, IL, Sept.
  1996.

\bibitem{cai:c:boolean-1}
{\sc J.~Cai}, {\em Probability one separation of the boolean hierarchy}, in
  Proceedings of the 4th Annual Symposium on Theoretical Aspects of Computer
  Science, Springer-Verlag {\it Lecture Notes in Computer Science \#247}, 1987,
  pp.~148--158.

\bibitem{cai-gun-har-hem-sew-wag-wec:j:bh1}
{\sc J.~Cai, T.~Gundermann, J.~Hartmanis, L.~Hemachandra, V.~Sewelson,
  K.~Wagner, and G.~Wechsung}, {\em The boolean hier\-archy {I}: {S}truc\-tural
  proper\-ties}, SIAM Jour\-nal on Com\-pu\-ting, 17 (1988), pp.~1232--1252.

\bibitem{cai-gun-har-hem-sew-wag-wec:j:bh2}
\leavevmode\vrule height 2pt depth -1.6pt width 23pt, {\em The boolean
  hierarchy {II}: Applications}, SIAM Journal on Computing, 18 (1989),
  pp.~95--111.

\bibitem{cai-hem:c:boolean}
{\sc J.~Cai and L.~Hemachandra}, {\em The boolean hierarchy: {H}ardware over
  {{N}{P}}}, in Proceedings of the 1st Structure in Complexity Theory
  Conference, Springer-Verlag {\it Lecture Notes in Computer Science \#223},
  June 1986, pp.~105--124.

\bibitem{cha:thesis:boolean}
{\sc R.~Chang}, {\em On the Structure of {NP} Computations under Boolean
  Operators}, PhD thesis, Cornell University, Ithaca, NY, 1991.

\bibitem{cha-kad:j:closer}
{\sc R.~Chang and J.~Kadin}, {\em The boolean hierarchy and the polynomial
  hierarchy: A closer connection}, SIAM Journal on Computing, 25 (1996),
  pp.~340--354.

\bibitem{gre-koe-reg-sch-tor:j:middle-bit}
{\sc F.~Green, J.~K{\"{o}}bler, K.~Regan, T.~Schwentick, and J.~Tor{\'{a}}n},
  {\em The power of the middle bit of a {\#}{P} function}, Journal of Computer
  and System Sciences, 50 (1995), pp.~456--467.

\bibitem{hem-hem-hem:jtoappear:downward-translation}
{\sc E.~Hemaspaandra, L.~Hemaspaandra, and H.~Hempel}, {\em A downward collapse
  within the polynomial hierarchy}, SIAM Journal on Computing.
\newblock To appear.

\bibitem{hem-hem-hem:jtoappear:query-order-survey}
\leavevmode\vrule height 2pt depth -1.6pt width 23pt, {\em An introduction to
  query order}, Bulletin of the EATCS.
\newblock To appear.

\bibitem{hem-hem-hem:jtoappear:sn-1tt-np-completeness}
\leavevmode\vrule height 2pt depth -1.6pt width 23pt, {\em R${}^{{\cal S}{\cal
  N}}_{1\hbox{-}tt}$({N}{P}) distinguishes robust many-one and {T}uring
  completeness}, Theory of Computing Systems.
\newblock To appear.

\bibitem{hem-hem-hem:ctoappear:query-order-ph}
\leavevmode\vrule height 2pt depth -1.6pt width 23pt, {\em Query order in the
  polynomial hierarchy}, in Proceedings of the
  11th Conference on Fundamentals of Computation Theory, Springer Verlag {\it
  Lecture Notes in Computer Science \#1279}, Sept. 1997.
\newblock To appear.

\bibitem{hem-hem-hem:t:translating-downwards}
\leavevmode\vrule height 2pt depth -1.6pt width 23pt, {\em Translating equality
  downwards}, Tech. Report TR-657, University of Rochester, Department of
  Computer Science, Rochester, NY, Apr. 1997.

\bibitem{hem-hem-wec:tOutBySICOMP:query-order-bh}
{\sc L.~Hemaspaandra, H.~Hempel, and G.~Wechsung}, {\em Query order and
  self-specifying machines}, Tech. Report TR-596, University of Rochester,
  Department of Computer Science, Rochester, NY, Oct. 1995.

\bibitem{hem-rot:j:boolean}
{\sc L.~Hemaspaandra and J.~Rothe}, {\em Unambiguous computation: {B}oolean
  hierarchies and sparse {T}uring-complete sets}, SIAM Journal on Computing, 26
  (1997), pp.~634--653.

\bibitem{her:c:transformation-monoid-acceptance}
{\sc U.~Hertrampf}, {\em Acceptance by transformation monoids (with an
  application to local self-reductions)}, in
  Proceedings of the 12th Annual IEEE Conference on Computational Complexity,
  IEEE Computer Society Press, June 1997, pp.~213--224.

\bibitem{kad:joutdatedbychangkadin:bh}
{\sc J.~Kadin}, {\em The polynomial time hierarchy collapses if the boolean
  hierarchy collapses}, SIAM Journal on Computing, 17 (1988), pp.~1263--1282.
\newblock Erratum appears in the same journal, 20(2):404.

\bibitem{koe-sch-wag:j:diff}
{\sc J.~K{\"{o}}bler, U.~Sch\"{o}ning, and K.~Wagner}, {\em The difference and
  truth-table hierarchies for {NP}}, RAIRO Theoretical Informatics and
  Applications, 21 (1987), pp.~419--435.

\bibitem{lad-lyn-sel:j:com}
{\sc R.~Ladner, N.~Lynch, and A.~Selman}, {\em A comparison of polynomial time
  reducibilities}, Theoretical Computer Science, 1 (1975), pp.~103--124.

\bibitem{pap:b:complexity}
{\sc C.~Papadimitriou}, {\em Computational Complexity}, Addison-Wesley, 1994.

\bibitem{wag:t:n-o-q-87version}
{\sc K.~Wagner}, {\em Number-of-query hierarchies}, Tech. Report 158,
  Universit\"{a}t Augsburg, Institut f\"{u}r Mathematik, Augsburg, Germany,
  Oct. 1987.

\bibitem{wag:t:n-o-q-89version}
\leavevmode\vrule height 2pt depth -1.6pt width 23pt, {\em Number-of-query
  hierarchies}, Tech. Report~4, Universit\"{a}t W\"urzburg, Institut f\"{u}r
  Informatik, W\"urzburg, Germany, Feb. 1989.

\bibitem{wag:j:bounded}
\leavevmode\vrule height 2pt depth -1.6pt width 23pt, {\em Bounded query
  classes}, SIAM Journal on Computing, 19 (1990), pp.~833--846.

\bibitem{wag:t:parallel-difference}
\leavevmode\vrule height 2pt depth -1.6pt width 23pt, {\em A note on parallel
  queries and the difference hierarchy}, Tech. Report 173, Universit\"at
  W\"urzburg, Institut f\"ur Informatik, W\"urzburg, Germany, June 1997.

\bibitem{wec:c:bh:ormaybe:wech:only:is:right}
{\sc G.~Wechsung}, {\em On the boolean closure of {NP}}, in Proceedings of the
  5th Conference on Fundamentals of Computation Theory, Springer-Verlag {\it
  Lecture Notes in Computer Science \#199}, 1985, pp.~485--493.
\newblock (An unpublished precursor of this paper was coauthored by K. Wagner).

\end{thebibliography}
\end{document}